\documentclass[12pt]{spieman}  
\usepackage{amsmath,amsfonts,amssymb}
\usepackage{graphicx}
\usepackage{setspace}
\usepackage{tocloft}
\usepackage{lineno}

\title{US National Gemini Office in the NOIRLab era}

\author[a,*]{Vinicius M. Placco}
\author[a]{Letizia Stanghellini}
\affil[a]{NSF’s NOIRLab, 950 N. Cherry Ave., Tucson, AZ 85719, USA}

\cftpagenumbersoff{figure}
\cftpagenumbersoff{table} 
\begin{document} 
\maketitle


\begin{abstract}
This article presents an overview of the US National Gemini Office (US NGO) and its role within the International Gemini Observatory user community. Throughout the years, the US NGO charter changed considerably to accommodate the evolving needs of astronomers and the observatory. The current landscape of observational astronomy requires effective communication between stakeholders and reliable/accessible data reduction tools and products, which minimize the time between data gathering and publication of scientific results. Because of that, the US NGO heavily invests in producing data reduction tutorials and cookbooks.
Recently, the US NGO started engaging with the Gemini user community through social media, and the results have been encouraging, increasing the observatory's visibility. The US NGO staff developed tools to assess whether the support provided to the user community is sufficient and effective, through website analytics and social media engagement numbers. These quantitative metrics serve as the baseline for internal reporting and directing efforts to new or current products.
In the era of the NSF's National Optical-Infrared Astronomy Research Laboratory (NOIRLab), the US NGO is well-positioned to be the liaison between the US user base and the Gemini Observatory. Furthermore, collaborations within NOIRLab programs, such as the Astro Data Lab and the Time Allocation Committee, enhance the US NGO outreach to attract users and develop new products. The future landscape laid out by the Astro 2020 report confirms the need to establish such synergies and provide more integrated user support services to the astronomical community at large.
\end{abstract}


{\noindent \footnotesize\textbf{*}Vinicius M. Placco,  \linkable{vinicius.placco@noirlab.edu}}

\begin{spacing}{1.12} 

\section{Introduction}
\label{sect:intro}  

The US National Gemini Office (US NGO) is the liaison between the International Gemini Observatory and the US astronomy community. The US NGO, as it exists today, was officially created in 1994 as the US Gemini Program (USGP). Since then, it morphed into the NOAO Gemini Science Center (NGSC) in 2003, and then the NOAO System Science Center (NSSC)'s System User Support (SUS) group in 2009. Finally, as of 2015, the SUS group became the US NGO. Over its almost 30 years of existence, the US NGO has changed and evolved based on the interactions with the users' community and the observatory, as recorded in a series of agreements with Gemini. For a complete historical overview of the US NGO, we refer the reader to the article by Hinkle et al.\cite{hinkle2018}. 

The US NGO is one of the several National Gemini Offices (NGOs) representing the Gemini International Partnership, which currently includes the United States, Canada, the Republic of Korea, Brazil, Argentina, and Chile. Gemini operates twin 8.1-meter diameter telescopes in Cerro Pachón, Chile (Gemini South, dedicated on January 18, 2002), and Maunakea, Hawai‘i, USA (Frederick C. Gillett Gemini North, dedicated on June 25, 1999). Partner countries are responsible for creating and maintaining their own NGOs, with the goal of supporting and advocating for their national scientific communities. Until 2019, the US NGO was an office within the National Optical Astronomy Observatory (NOAO), and it is now part of the Community Science and Data Center (CSDC) program at the NSF's National Optical-Infrared Astronomy Research Laboratory (NOIRLab). 

The first charter of the US NGO included user support and advocacy for the Gemini telescopes in the US community. Until 2014, the US NGO, similar to other NGOs, had actively supported its users in the {\it Phase II}\footnote{The life cycle of Gemini Programs is defined as {\it Phase I} (proposal for telescope time and time allocation), {\it Phase II} (program preparation, observations, quality control, and archiving), and {\it Phase III} (data reduction and analysis). Please refer to \linkable{https://www.gemini.edu/observing/start-here} for further information.} of the observing cycle -- from proposal allocation through the acquisition of the observations. After 2014, given the large number of accepted US programs (on average, 97 per semester on both telescopes) and the science instrument expertise at Gemini, it became clear that the observatory staff should directly support the US {\it Phase II} proposals. 
In exchange for {\it Phase II} support, the US NGO was tasked to develop, in collaboration with the Science User Support Department (SUSD) at Gemini, a semesterly {\it exchange work package}, which contains a set of activities determined and agreed upon by CSDC and Gemini. The goal of these work packages is to benefit the entire international Gemini community, and they have historically leaned towards {\it Phase III} support --- from data acquisition to science, most notably through tutorials and data reduction cookbooks --- but not exclusively. For example, the deployment of the updates and the installation of Phoenix at Gemini South in 2016 were also part of the exchange package between the US NGO and Gemini.

The role of the US NGO was discussed and re-staged when NOAO transitioned into NSF's NOIRLab in 2019.  The US NGO had the opportunity to renew itself and expand its scope.  It continues to act in its double role of advocacy for the US Gemini community and focal point for {\it Phase III} user support for the whole community and developed an active role in engaging the broad Gemini community, including a lively social media presence. The US NGO, now a group within CSDC at NSF's NOIRLab, shares a high-level management structure with Gemini. Nonetheless, it still maintains its independence in terms of governance and funding lines. The US NGO interacts with Gemini and the other NGOs through activities such as the Gemini Operations Working Group meeting -- which is advisory to the Gemini Director on the use and scheduling of the Gemini telescopes -- and the joint Gemini-NGOs meetings. US NGO representatives also participate in the Gemini-specific merging section of the NOIRLab Telescope Allocation Committee (TAC), providing occasional Gemini-specific expertise in the final proposal raking for the US partner.
 
This article showcases the US NGO activities, focusing on the most recent developments. Section~\ref{charter} outlines the US NGO's current charter, including the services provided for both the US and the entire Gemini community. Section~\ref{engagement} illustrates the current efforts related to community engagement and user support, followed by the synergies with other NOIRLab programs in Section~\ref{synergies}. We provide our view of the future of the US NGO in Section~\ref{future}.

\section{Current Charter}
\label{charter}

The US NGO is one of the partner's National Offices that shares with Gemini the responsibilities to support the users in the phases of the astronomical observing cycle. In particular, the US NGO is responsible for the partner {\it Phase I} -- proposal preparation -- and {\it Phase III} -- post-observation data processing. Such support includes responding to helpdesk queries in specific categories for all Gemini partners (see details in Section~\ref{gemini_colab}); Gemini staff executes the observations and provides on-site support for users' requests, also through specific categories in the Gemini helpdesk. The difference between the US NGO and other partner countries' NGOs is that in the US case, Gemini staff (rather than the US NGO staff) is responsible to check and validate the US {\it Phase II} programs that will populate the Gemini observing queue, while other partners do so in collaboration with the Gemini experts. In turn, the US NGO provides products for the broad Gemini community through the {\it exchange work package}, as mentioned in the Introduction\footnote{for further details, see also ``Introductory material for Gemini Governance committees members'', available at \linkable{https://www.gemini.edu/science/}.}.

Given the framework of the slightly different agreement with Gemini, the US NGO activities can be divided into (1) support and advocacy of US users and (2) community engagement and support for the international Gemini community. For the US Gemini users, the main goal of the US NGO is to ensure that the US PIs have the support they need to write successful proposals and analyze their data, and to minimize the time between data acquisition and scientific publications. The US NGO staff achieves this level of support with an array of activities, described in Section~\ref{engagement}, which are selected based on institutional agreements/commitments and internal/community feedback. The US NGO answers most Gemini helpdesk queries for the US partner and selected categories for the other partners, develops data reduction cookbooks and tutorials, informs the US community about the Gemini initiatives, deadlines, instrumentation, and other issues, and holds specific science and technology meetings within the AAS meeting framework\footnote{\linkable{https://noirlab.edu/science/programs/csdc/usngo/workshops}.}. 

Over the years, the uniqueness of the agreement between Gemini and NOAO/NOIRLab/CSDC has also been reflected in the evolution of staffing efforts within the US NGO. Until 2014, there were up to twelve scientific and four technical/administrative staff (mostly shared with other NOAO departments) working for the US NGO. Once the US {\it Phase II} work became Gemini's responsibility circa 2014, the US NGO staff were approximately 3 FTE (Full-time equivalent), decreasing to 1.5 FTE in 2017. Since 2020, the US NGO is staffed at a 2.5 FTE level.

Since the beginning of the Gemini Observatory science operations in the early 2000s, the US has been the majority partner with roughly 2/3 of the available observing time in both telescopes every semester. Between the 2005A and 2023A semesters, the US TAC has received, on average, 209 proposals (127 for Gemini North and 82 for Gemini South) requesting 2894 hours (1765 for Gemini North and 1129 for Gemini South) of observations per semester for the Gemini US partner. The US NGO staff works alongside Gemini and the NOIRLab TAC to ensure that the scheduling and execution of the US-approved science programs are within the users' requests.


The US NGO keeps open communications with Gemini staff and users to identify potential opportunities. For example, we may advertise modes of operation, instrument updates, schedule changes, and new software releases. By broadcasting the various Gemini modes of observation and instrumentation, the US NGO informs the community about what is available and attracts new users to Gemini. In addition, the US NGO staff aims to continually engage with the user community to gather input on what may be missing and feedback on the products and services currently offered.

The US NGO agrees on semesterly with Gemini on services to support the broad community. These may include data reduction support and the development of tutorials and cookbooks for the facility instruments, providing expertise to address helpdesk questions from other partners, among others. In addition, the US NGO head serves as an ex-officio member of the Users' Committee for Gemini (UCG), representing all the National Gemini Offices.

Similarly to all the National Gemini Offices, {\it the US NGO is not a part of the Gemini Observatory}. As stated above, the US NGO is a group within CSDC at NOIRLab. CSDC and Gemini are separately managed programs within NOIRLab, with separate funding and leadership lines. Nonetheless, they share a matrix framework for managing the scientific and engineering staff in Gemini and CSDC, facilitating collaborative work between the US NGO members and the Gemini staff.

\section{Community Engagement and User Support}
\label{engagement}

\subsection{Web Portal}

\begin{figure}[!ht]
\begin{center}
\includegraphics[scale=0.525]{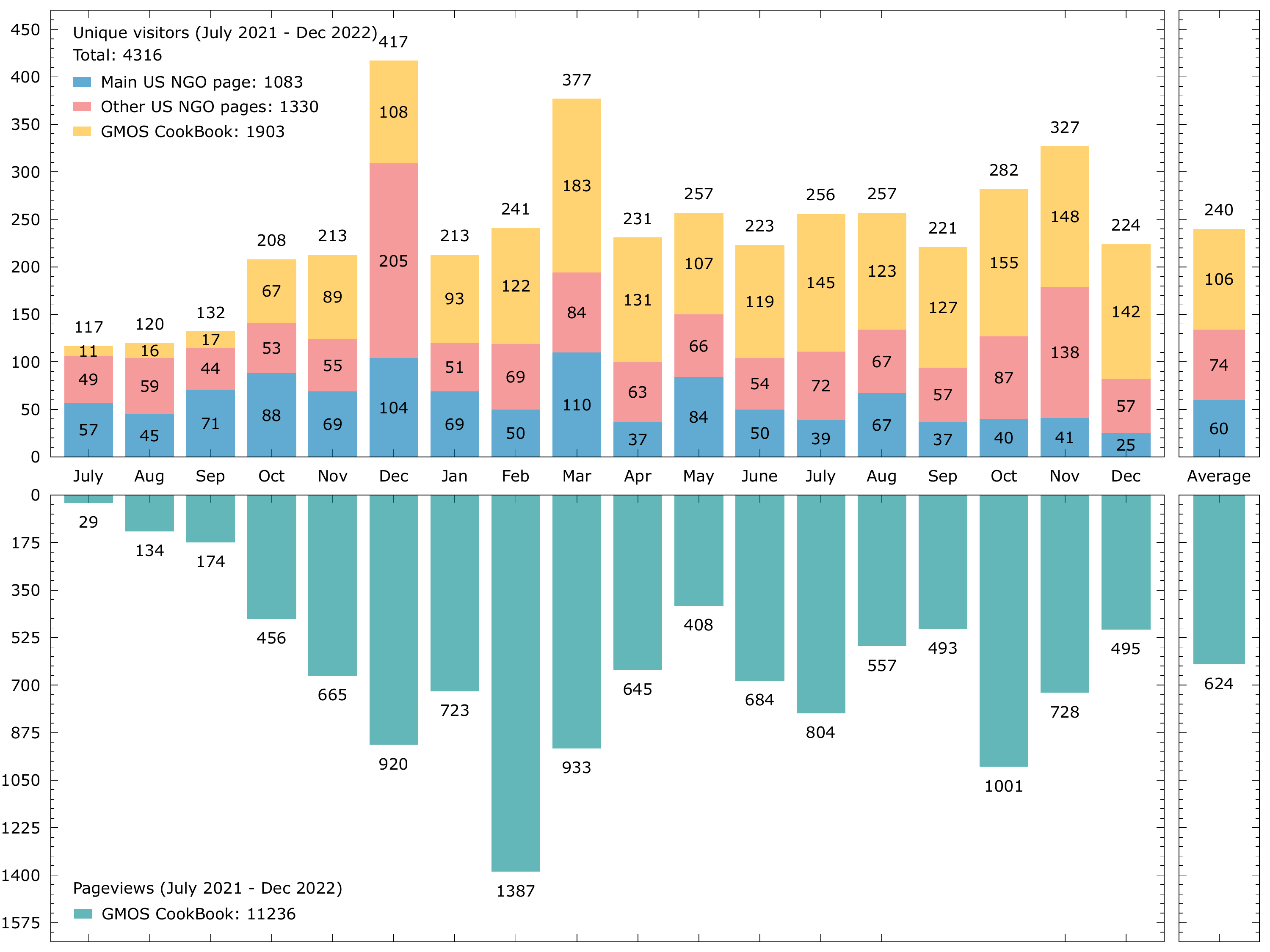}
\end{center}
\caption 
{ \label{web_analytics} Upper panel: Number of unique visitors per month for the US NGO main web page, other US NGO pages, and the GMOS Cookbook, in the period of July 2021 to December 2022. Lower panel: Number of views for all the GMOS Cookbook pages in the same period. The average values per month are shown in the rightmost panels.} 
\end{figure}

The US NGO web portal (\linkable{https://noirlab.edu/science/programs/csdc/usngo}) contains 
important resources for the entire Gemini community. There are curated pages with data reduction tutorials and presentations for current and retired instruments, as well as a collection of links with quick access to data analysis resources and proposal preparation pages. The web page also highlights relevant operations updates from Gemini, including scheduled shutdowns, instrument availability, and software updates. The website has been recently updated and launched in tandem with the new NOIRLab Science website (\linkable{https://noirlab.edu/science/}). The upper panel of Figure~\ref{web_analytics} illustrates the importance of our web pages to the Gemini community, showing the number of unique visitors between July 2021 and December 2022. The rightmost upper panel shows the 18-month average for the same period, with 60 unique visitors to the main web page per month and 74 for the other pages within the website.

\subsection{Data reduction cookbooks and tutorials}


The two Gemini Multi-Object Spectrographs (GMOS\cite{davies1997,gimeno2016}) were the first instruments to be delivered at each telescope (GMOS-N in July 2001 and GMOS-S in December 2002), providing imaging and spectroscopy modes since the start of the Gemini science operations. Both spectrographs remain among the most requested observing resources for the entire Gemini community\footnote{See \linkable{https://www.gemini.edu/instrumentation/gmos} for further information.}.

Based upon community feedback, the US NGO created the GMOS Data Reduction Cookbook in 2016 (\linkable{https://noirlab.edu/science/programs/csdc/usngo/gmos-cookbook/}). This is an extensive tutorial that thoroughly describes all the GMOS observing modes and provides step-by-step instructions on how to reduce archival data using the Gemini-specific IRAF/PyRAF packages\footnote{IRAF (Image Reduction and Analysis Facility) was distributed by the National Optical Astronomy Observatory, which was managed by the Association of Universities for Research in Astronomy (AURA) under a cooperative agreement with the National Science Foundation\cite{tody1986,tody1993}.}, from raw frames to science-quality spectra/images. It also contains more general useful resources for users and supplementary materials of relevance. The US NGO also created and maintains the GMOS Observation Planner (\linkable{https://noirlab.edu/science/programs/csdc/usngo/gmos-obsplan/}), which helps users plan and prepare GMOS observations.

\begin{figure}[!ht]
\begin{center}
\includegraphics[scale=0.4]{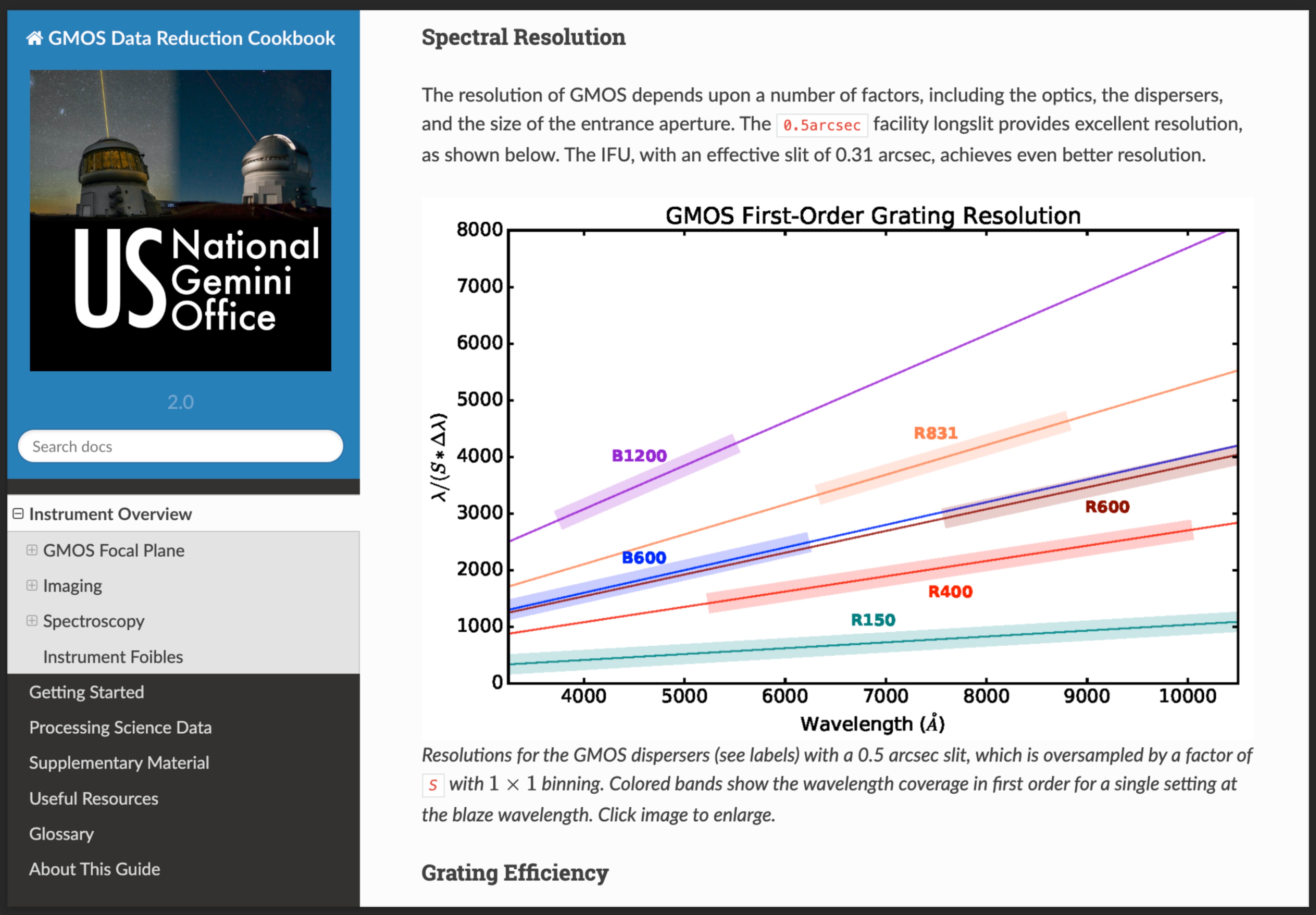}
\end{center}
\caption 
{ \label{cookbook_snap} Snapshot of the GMOS Cookbook, hosted at \linkable{https://noirlab.edu/science/programs/csdc/usngo/gmos-cookbook/}} 
\end{figure}

The GMOS Cookbook has been recently upgraded (v2.0) and tested with the latest software released by the observatory\cite{merino2022}. Figure~\ref{cookbook_snap} shows a snapshot of the GMOS instrument overview page within the Cookbook. This product serves the entire Gemini user community worldwide. The Cookbook receives substantial attention from the user community (see the upper panel of Figure~\ref{web_analytics}). Furthermore, the lower panel of Figure~\ref{web_analytics} shows the monthly Cookbook page views for the past 16 months, with a total of over 10,000. Since its launch, v2.0 of the Cookbook has had 101 unique visitors and 626 page views per month on average.

\begin{figure}[!ht]
\begin{center}
\includegraphics[scale=0.32]{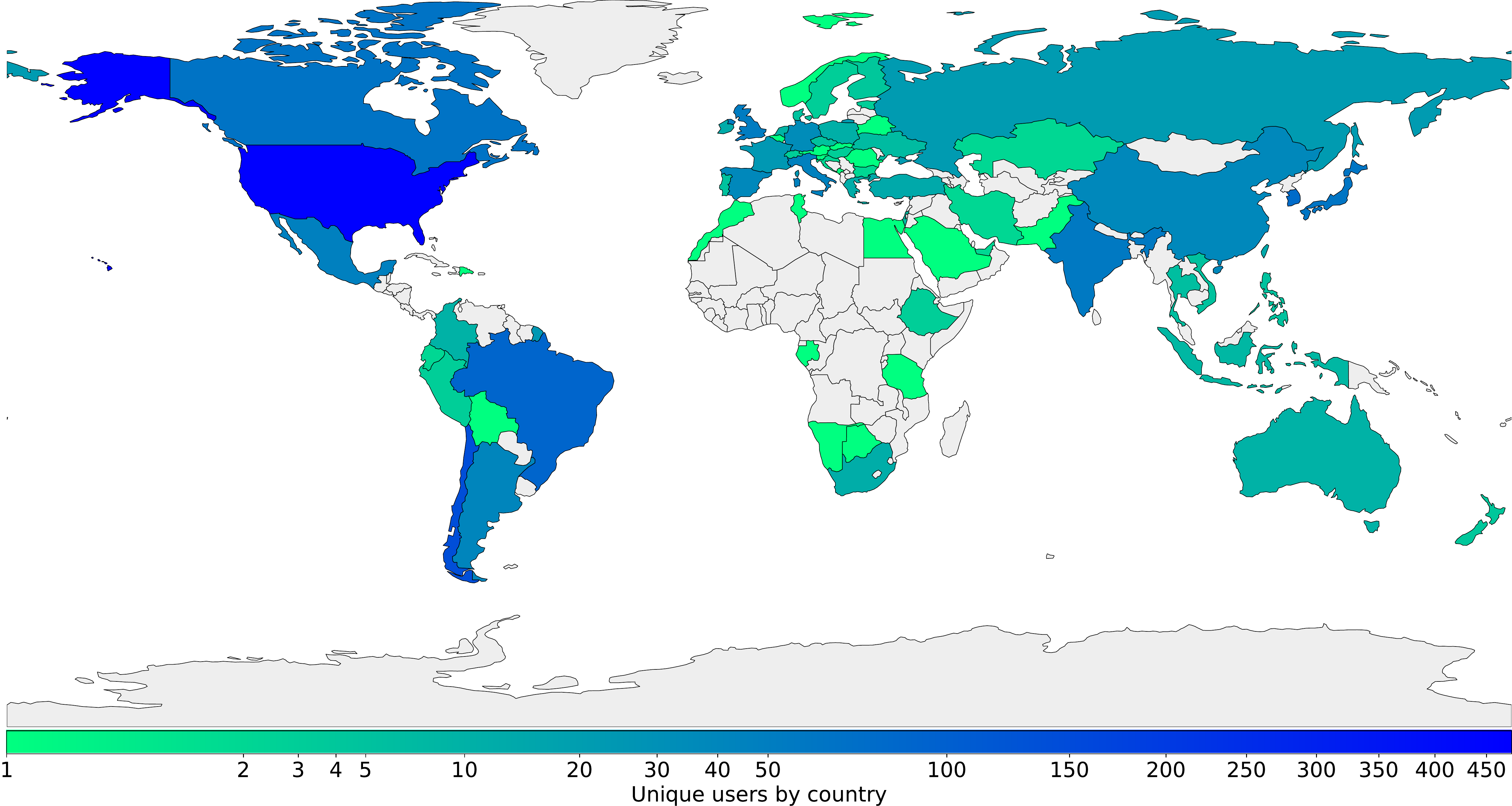}
\end{center}
\caption 
{ \label{cookbook_map} GMOS Cookbook usage across the globe between July 2021 and December 2022. The map is color-coded by the number of unique users in each of the 69 countries with reported data by Google Analytics. Data were retrieved on February 1, 2023.} 
\end{figure}

Figure~\ref{cookbook_map} further highlights the outreach of the GMOS Cookbook. Based on site statistics gathered between July 2021 and December 2022, the Cookbook has been accessed by users in 69 different countries. Moreover, all the current Gemini partner countries are among the top ten in terms of unique users (USA: 476, Chile: 140, Brazil: 89, South Korea: 79, Canada: 65, and Argentina: 42). These numbers further emphasize the importance of maintaining such resources for the entire Gemini community. 

\begin{figure}[!ht]
\begin{center}
\includegraphics[scale=0.37]{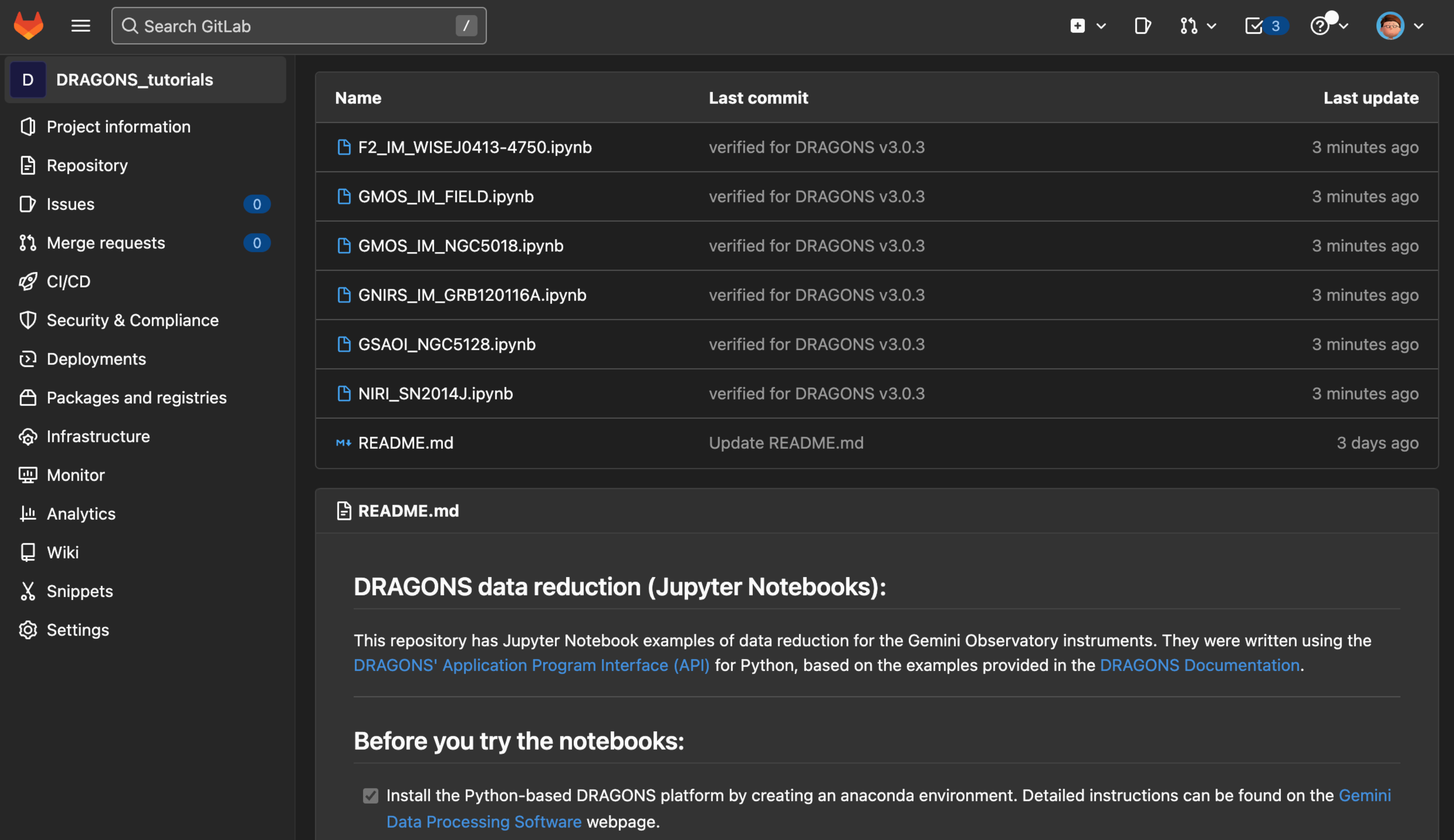}
\end{center}
\caption 
{ \label{gitlab_snap} Snapshot of the GitLab repository page, containing the DRAGONS Jupyter Notebooks with data reduction tutorials (\linkable{https://gitlab.com/nsf-noirlab/csdc/usngo/DRAGONS\_tutorials})} 
\end{figure}

In support of the users community, the US NGO created data reduction tutorials in Python, in the form of Jupyter Notebooks (\linkable{https://gitlab.com/nsf-noirlab/csdc/usngo/DRAGONS\_tutorials}). Figure~\ref{gitlab_snap} shows a snapshot of the \texttt{README} page that contains the current list of tutorials. These cover all the instruments and modes currently available for DRAGONS\cite{dragons} (Data Reduction for Astronomy from Gemini Observatory North and South), which is the Python-based software suite being developed by Gemini. The notebooks contain information on how to download the raw data directly from the Gemini Observatory Archive (\linkable{https://archive.gemini.edu/}) and all the processing steps to generate a final science-ready product.

\subsection{Social media engagement - Twitter}

In September 2020 the US NGO created a Twitter account (\linkable{https://twitter.com/usngo}) with the goals of (i) providing another channel of communication between the users and the US NGO staff; (ii) promoting Gemini capabilities, sharing opportunities, and operations updates; (iii) engaging with the astronomical community at large by developing products related to Gemini publications and data products (see details below). The response from the community has been extremely positive, increasing the observatory's visibility and bringing new users to Gemini.

\begin{figure}[!ht]
\begin{center}
\includegraphics[scale=0.54]{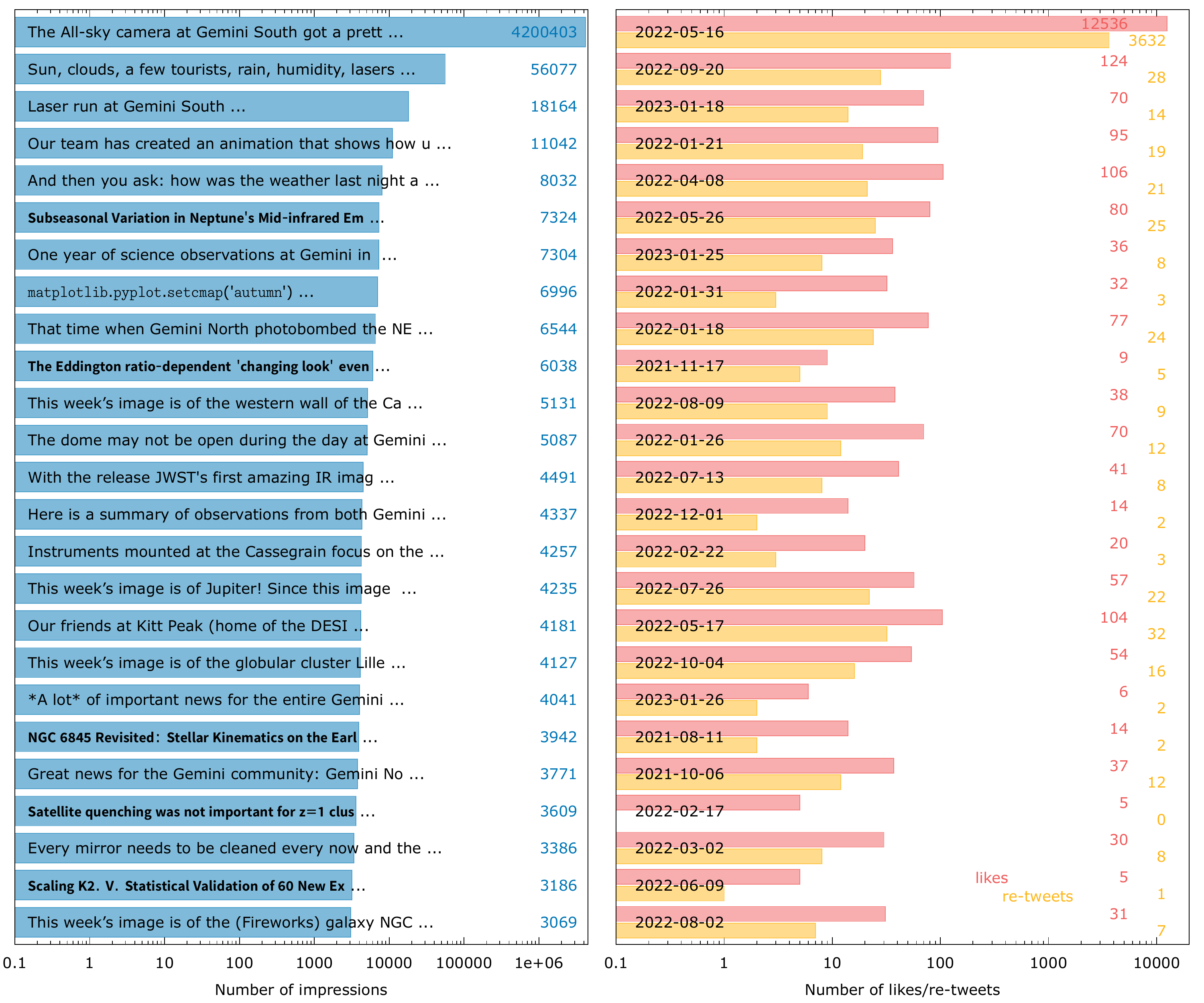}
\end{center}
\caption 
{ \label{twitter} Left panel: Number of impressions (times a Tweet was seen on Twitter - numbers in blue) for the top 25 posts on the US NGO account since September 2020. An excerpt of the Tweet is given in each bar and the texts in bold font show paper highlights. Right panel: Number of likes (red) and re-tweets (yellow) for the Tweets in the left panel. Also shown are the publication dates. Data were retrieved on February 1, 2023.} 
\end{figure}

The results of this effort can be appreciated in Figure~\ref{twitter}. All the statistics shown here were retrieved using the Twitter API (Application Programming Interface). The left panel shows the number of impressions (times a Tweet was seen on Twitter - numbers in blue) for the top 25 posts on the US NGO account since September 2020. An excerpt of the Tweet is given in each bar. Tweets with text in boldface refer to the ``publication of the week'' series, which highlights a refereed publication with Gemini data and US participation. As of February 1, 2023, this series has featured 101 articles, gathering 113,260 impressions combined.

Most publications on Twitter receive, on average, from several hundred to a few thousand impressions. Some notable exceptions include the 2022 Lunar Eclipse at Gemini South\footnote{\linkable{https://twitter.com/usngo/status/1526336287749263362}} (with over 4.2 million impressions), a time-lapse video at Gemini North\footnote{\linkable{https://twitter.com/usngo/status/1572245161882832896}} (with over 56,000 impressions), an animation with all the Gemini science frames observed in 2021\footnote{\linkable{https://twitter.com/usngo/status/1484545341328224261}} (with about 11,000 impressions), and a refereed article on observations of Neptune\footnote{\linkable{https://twitter.com/usngo/status/1529924220360089600}} (with over 7,300 impressions). The right panel of Figure~\ref{twitter} shows the publication date (black), the number of likes (red), and the number of re-tweets (yellow) for the Tweets in the left panel. In July 2022, the US NGO started the ``Image of the week'' series, with threads that briefly describe an image from the Gemini archive and highlight refereed articles that study the objects associated with the images. The response has also been positive, with 5 threads among the top-25 publications shown in Figure~\ref{twitter}.

\begin{figure}[!ht]
\begin{center}
\includegraphics[scale=0.45]{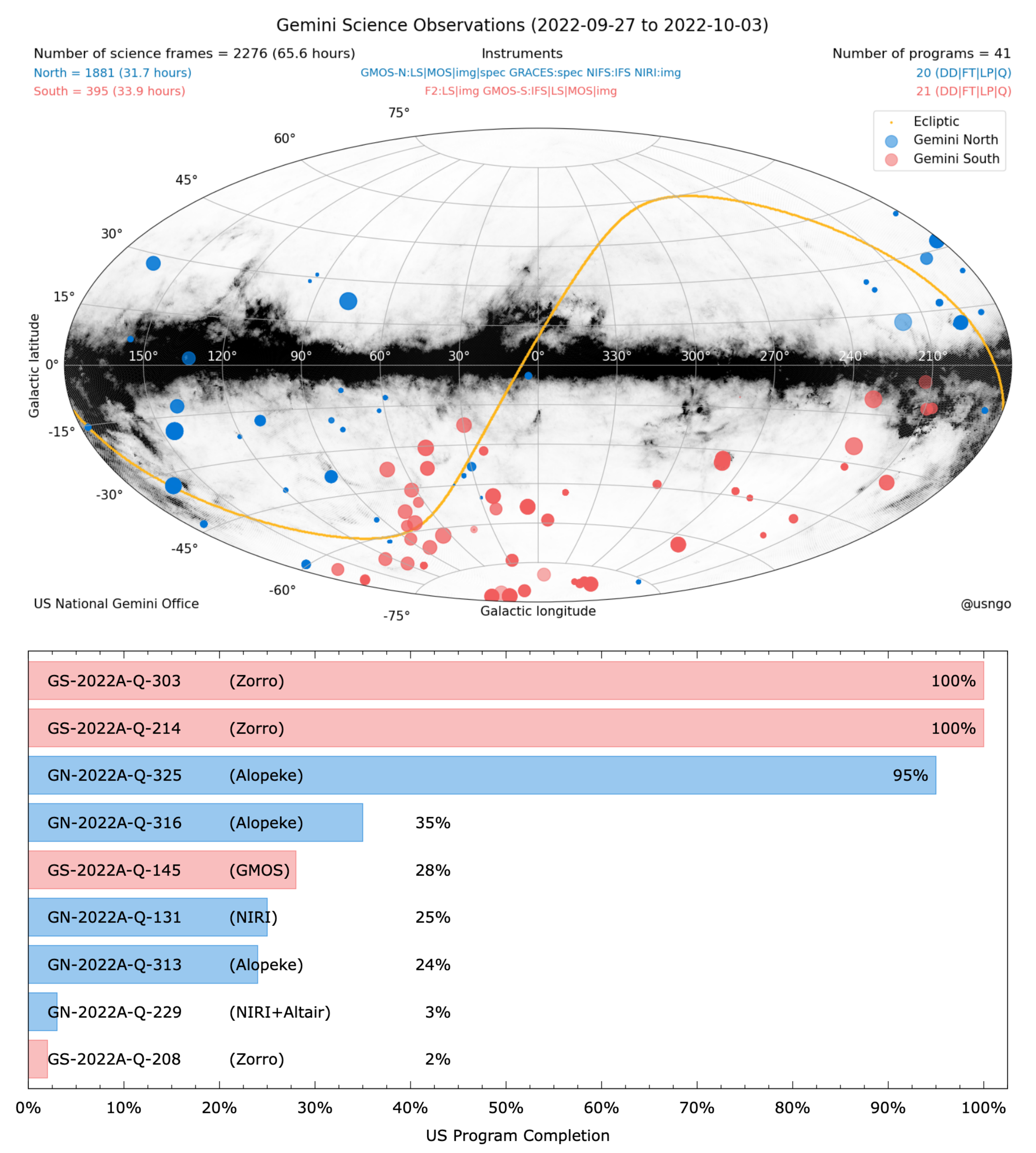}
\end{center}
\caption 
{ \label{obs_summary} Top: Example of weekly program completion (September 27 to October 3, 2022). Bottom: Example of daily US program completion (May 9, 2022). See text for details.} 
\end{figure} 

In addition, the US NGO has developed a number of ``Twitter bots'', which autonomously publish information to the Twitter account. All the information used is taken either from the public Gemini web pages or from the Gemini archive. Two such examples are shown in Figure~\ref{obs_summary}. The top panel shows a weekly summary, which includes the galactic coordinates for all the science frames observed on the previous week, as well as the number/type of programs, total hours on target, and instruments/modes used. 
The bottom panel shows the daily program completion summary. For each US queue program with observations taken on the previous night, the summary shows the program number, the instrument used, and the completed fraction of the allocated time, color-coded by the telescope used. Typical time allocations for Gemini US queue programs can range from about 30 minutes to tens of hours. As an example, program GS-2022A-Q-214 (second bar from the top in the bottom panel of Figure~\ref{obs_summary}) had an allocated time of 1.10 hours and was fully executed on May 9, 2022.
These summaries help create engagement with the community by showing how effectively both telescopes are operating on a daily/weekly basis.

As of February 1st, 2023, the US NGO's Twitter account had 1,320 followers.


\subsection{Publication metrics and citation tracking}

\begin{figure}[!ht]
\begin{center}
\includegraphics[scale=0.65]{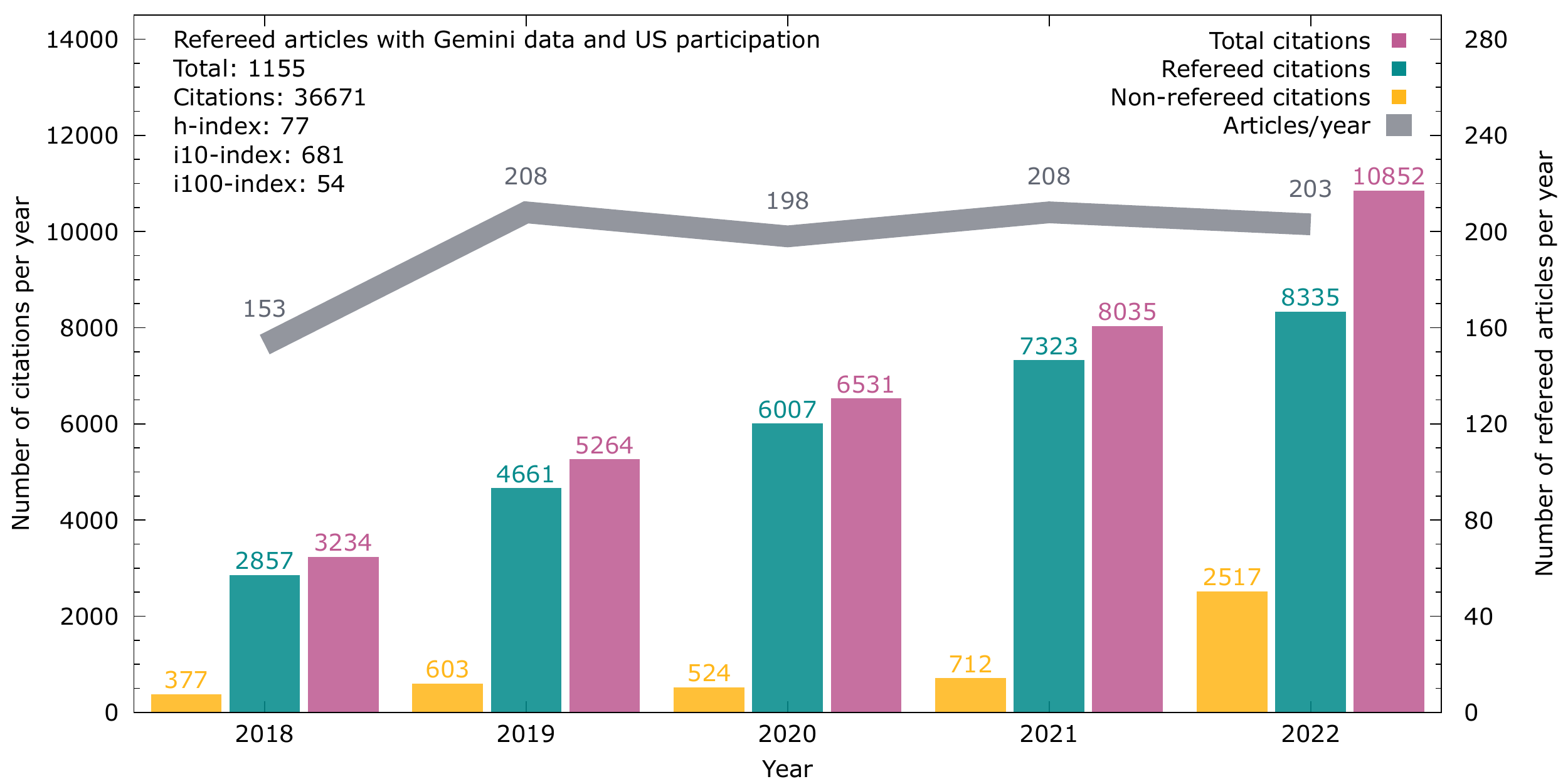}
\end{center}
\caption 
{ \label{pub_metrics} Publication metrics using the ADS API. Data pertains to refereed articles with Gemini data and at least one co-author with a US affiliation. The bars show the citations gathered in the 2018-2022 period and the line shows the number of published articles in each year. Publication metrics are given in the top left (see text for further details. Data were retrieved on February 1, 2023.} 
\end{figure}

The US NGO tracks publication metrics related to refereed astronomy articles with Gemini data and US participation. The reports are generated using the SAO/NASA Astrophysics Data System API (\linkable{https://ui.adsabs.harvard.edu/help/api/}).
Figure~\ref{pub_metrics} shows the evolution of citation counts per year (both refereed and non-refereed - colored bars) and the number of refereed articles (cyan line) between January 2018 and December 2022. The data were retrieved from the ADS bibliographic group \texttt{[bibgroup:gemini]} and only include articles with at least one author with a US affiliation. Also shown (top left) are the total number of articles, citations, and h-/i10-/i100- indices. 

The number of publications has remained steady between 2019 and 2022, despite the extended period that both Gemini telescopes remained closed due to the COVID-19 pandemic. This can be partially attributed to the use of archival data, while a potential decrease in publications for 2023 may happen as a delayed effect of the closures. The indices also demonstrate the growing scientific impact of publications with Gemini data. Within this period, 1155 refereed publications collected over 36,600 citations with an h-index of 77 (77 articles with at least 77 citations), including 54 articles with over 100 citations.

\subsection{Info-graphics}

\begin{figure}[!ht]
\begin{center}
\includegraphics[scale=0.34]{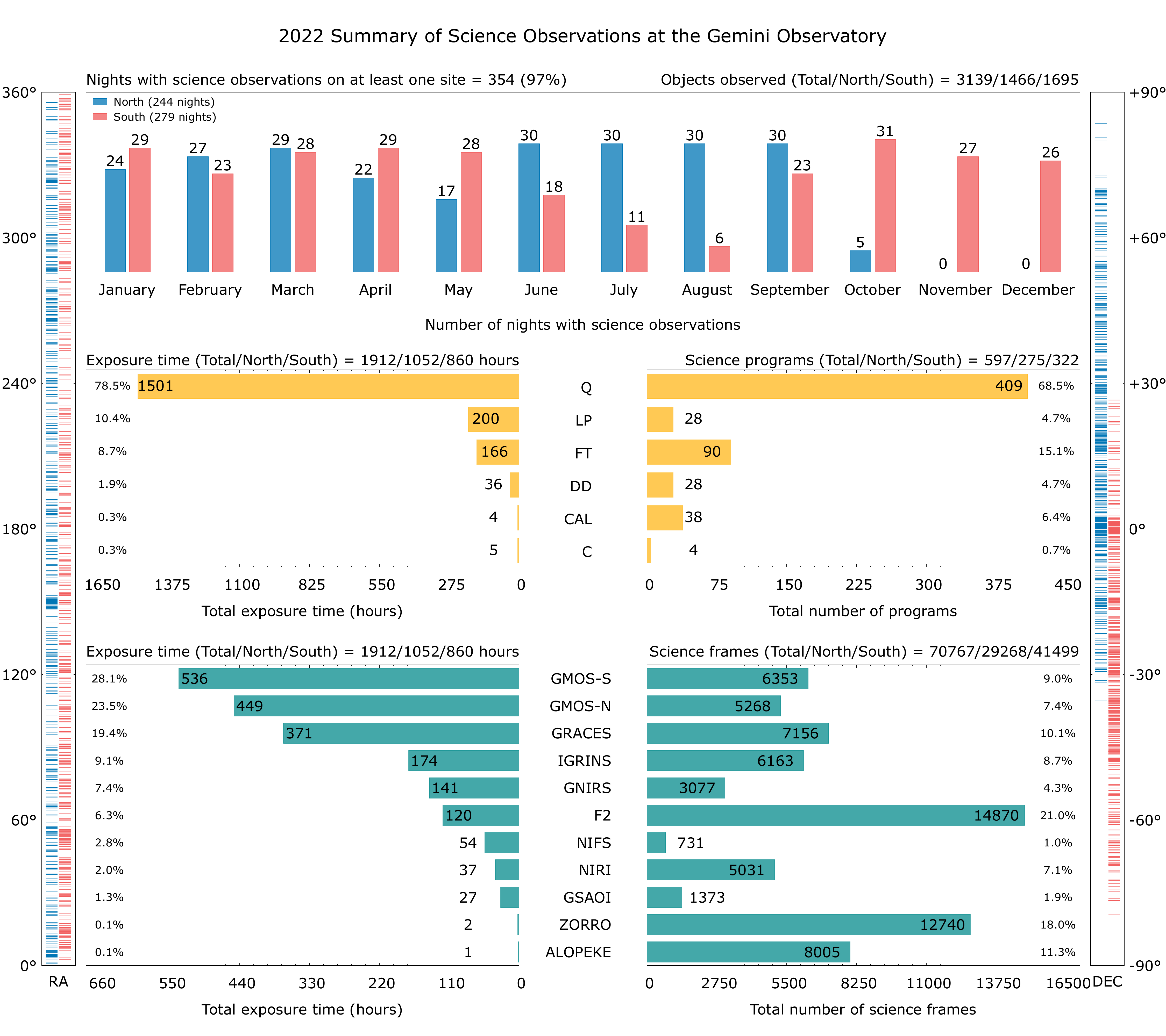}
\end{center}
\caption 
{\label{infographic} Summary of science observations collected by both Gemini telescopes in 2022. Public data retrieved from the Gemini Observatory Archive. See text for details.} 
\end{figure}

Another approach to engage with the user community and showcase the efficiency and flexibility of the Gemini telescopes is through the use of info-graphics. These are produced with data gathered through the Gemini Observatory Archive API\footnote{\linkable{https://archive.gemini.edu/help/api.html}}. Figure~\ref{infographic} shows an example of such a product, which is featured on social media and the US NGO website\footnote{\linkable{https://noirlab.edu/science/programs/csdc/usngo/gemini2022}}. It shows a summary of all science frames collected in 2022 at both telescopes, summarized by program type, instrument, observing date, and coordinates.

From the figure, science data were acquired by at least one Gemini telescope on 354 nights (97\%). Two notable gaps in the observations relate to the scheduled and extended shutdowns (August 1 - 25 for Gemini South and starting October 10 for Gemini North) and science observations were performed on at least 20 nights per month at each telescope for most of the year. The majority of the on-target science time (78.5\%) was spent observing the regular queue (Q), with Fast Turnaround (FT), Large and Long Programs (LP), and Director's Discretionary (DD) time combined for 21\% of the time. The remaining programs were classical (C) and calibration (CAL). Most of the science data were acquired with GMOS-N/S (52\% of the time, or 985 hours), followed by GRACES\footnote{Gemini Remote Access to CFHT ESPaDOnS Spectrograph. See \linkable{https://www.gemini.edu/instrumentation/graces} for further details.} and the Near-IR instruments. For the two speckle imaging instruments, `Alopeke and Zorro\footnote{\linkable{https://www.gemini.edu/instrumentation/alopeke-zorro}}, each frame counted by the Gemini Observatory Archive is a cube with one thousand or more images, so the real number of frames and exposure time are higher by at least a factor of 1000. The side panels show the RA and DEC distributions of the science targets. The coverage for right ascension is mostly complete, and there is a substantial overlap in declination close to the celestial equator.

Given the flexibility and robustness of the Gemini Observatory Archive, as evidenced by Figure~\ref{infographic}, there are still a number of opportunities to develop similar visualizations and statistics, which can benefit not only the current users but can also further inform the astronomical community at large and help bring new users to the Gemini facilities.

\section{Synergies with other NOIRLab programs}
\label{synergies}

\subsection{Gemini Observatory}
\label{gemini_colab}


The US NGO staff constantly interacts with the Gemini staff on various fronts. Notably, we have a close collaboration with the Science User Support Department (SUSD). The common thread between the US NGO and the SUSD is data reduction. The US NGO staff members work closely with Gemini colleagues in testing new software, performing science verification of new data reduction modes (using archival data) to be offered to the community, and also in common data reduction issues that may arise during interactions with users through the helpdesk system.
For the external Gemini helpdesk system, the US NGO staff members address all Tier 1\footnote{Tier 1 refers to the first interaction with the user. After a triage process, the ticket can be addressed still in Tier 1 or it can be escalated to Tiers 2 and 3, depending on the complexity of the issue raised.} tickets for the US partner\footnote{At this time, tickets submitted to the DRAGONS category are still handled by Gemini; in the near future, the US NGO will also respond to those.}, as well as Tier 2 for selected categories (GMOS and GMOS Cookbook), which include all other Gemini partners. The US NGO endeavors to have a first interaction with the user, either with a possible resolution of asking for further clarification, within one or two business days. Between 2018 and 2022, the US NGO has received an average of 44 helpdesk tickets per semester, with a success rate of resolving issues of about 95\%. Over the past four semesters, the median time between ticket submission and first interaction with the user has been 1.2 business days.

\subsection{Community Science and Data Center}


The US NGO collaborates with the NOIRLab TAC members, also within CSDC, who manage and handle all US-submitted Gemini proposals. The US NGO scientists are on-call during the US TAC process to answer questions about Gemini proposals, including technical reviews as requested by the TAC. The US NGO Head participates in the Gemini merging part of the US TAC, which consists of the merging of all the US Gemini proposals by rank. There is also occasional communication with proposal Principal Investigators for special queries arising at the Gemini merging TAC.
In addition, the US NGO also collaborates with the NOIRLab Astro Data Lab (\linkable{https://datalab.noirlab.edu/}), which provides an open-access/open-data science platform for big-data astronomy, enabling the efficient exploration and analysis of very large data sets. One project that is currently under development is the use of the Astro Data Lab Science Platform to remotely access, reduce, and analyze Gemini public and proprietary data. In this context, users would not have to install software locally or download any data. The latest software released by the observatory would be provided within a dedicated python environment where users could have their own workflows written as Jupyter Notebooks. This would allow for the use of the Gemini Observatory Archive and DRAGONS APIs, as well as other packages for data analysis.


\section{The future of the US NGO - An integrated vision in the NOIRLab era}
\label{future}

Throughout its almost 30 years of existence, the US NGO has taken several different roles and responsibilities within the International Gemini Observatory, NOAO, and NOIRLab. Nevertheless, its main focus has always been supporting the US users to ensure that the most optimal use of the observing resources is made and the best scientific results are produced as the outcome. Moreover, as stated by Hinkle et al., the US NGO has an essential role in protecting open access to 8-meter class facilities and advocating for both the US community and the Gemini Observatory. User engagement has greatly evolved since the early days of NOAO and Gemini, and the observatory staff, as stewards of the open-access astronomical facilities, must also adapt to this new paradigm. The US NGO initiative to engage with the Gemini user community through social media has proven so far to be an effective tool to increase the observatory's visibility and outreach. The US NGO staff endeavors to proactively seek feedback from the user community and identify further opportunities to develop new products.

Exciting times are ahead with the upcoming Gemini new suite of instruments (GHOST, Scorpio, IGRINS-2, GIRMOS, GPI 2.0) and the possible synergies with the Rubin Observatory's Legacy Survey of Space and Time. An investment must be made in robust and open-source data reduction pipelines, which are key for future success. The Astro 2020 report highlighted the need to support data archives and curation for the exponentially growing datasets from ground- and space-based facilities, and the importance of offering appropriate infrastructure and user support for data exploration and scientific discovery.

In this interim, the US NGO can be at the forefront for user support in the Gemini community, by establishing collaborations within NOIRLab and beyond and focusing on {\it Phase III} support. By working together with Gemini and CSDC, the US NGO will strive to continue minimizing the time between observations taken at the telescopes and maximizing the scientific outcome. The US NGO staff will continue to encourage and foster collaboration within NOIRLab and also with the other NGOs to the benefit of the entire astronomical community.

\subsection*{Disclosures}

The authors have no relevant financial interests in the manuscript and no other potential conflicts of interest to disclose.

\subsection* {Acknowledgments}

The work of V.M.P. and L.S. is supported by NOIRLab, which is managed by the Association of Universities for Research in Astronomy (AURA) under a cooperative agreement with the National Science Foundation. We thank Brian Merino for his important contributions to the US NGO products described in this paper. We thank Sharon Hunt for useful discussions and feedback regarding the publication metrics and reports generated by the US NGO.

%


\bibliography{report}   
\bibliographystyle{spiejour}   


\end{spacing}
\end{document}